\documentclass[10pt,letterpaper]{article}
\usepackage{opex3}
\usepackage{cite}

\begin{document}

%%%%%%%%%%%%%%%%%% title page information %%%%%%%%%%%%%%%%%%
\title{Optical transmittance degradation in tapered fibers}

\author{Masazumi Fujiwara$^{1,2}$, Kiyota Toubaru$^{1,2}$ and Shigeki Takeuchi$^{1,2,*}$}

\address{$^1$Research Institute for Electronic Science, Hokkaido University, Sapporo, Hokkaido 001-0021, Japan
\\
$^2$The Institute of Scientific and Industrial Research, Osaka University, Ibaraki, Osaka 567-0047, Japan}

\email{takeuchi@es.hokudai.ac.jp} %% email address is required



\begin{abstract}
We investigated the cause of optical transmittance degradation in tapered fibers. 
Degradation commences immediately after fabrication 
and it eventually reduces the transmittance to almost zero. 
It is a major problem that limits applications of tapered fibers. 
We systematically investigated the effect of the dust-particle density and the humidity on the degradation dynamics. 
The results clearly show that the degradation is mostly due to dust particles and that it is not related to the humidity.
In a dust free environment it is possible to preserve the transmittance with a degradation of less than the noise ($\pm$0.02) over 1 week.
\end{abstract}

 % REPLACE WITH CORRECT OCIS CODES FOR YOUR ARTICLE

%%%%%%%%%%%%%%%%%%%%%%% References %%%%%%%%%%%%%%%%%%%%%%%%%

%%%%%%%%%%%%%%%%%%%%%%%%%%  body  %%%%%%%%%%%%%%%%%%%%%%%%%%
\section{Introduction}
Tapered optical fibers are very powerful tool for various optical studies from quantum physics to biological sensing.
Their internal guided single modes have very high light coupling efficiencies ($> 90 \%$) with ultrahigh-Q optical microresonators.
This ideal cavity system can be used to investigate fundamental topics in cavity quantum electrodynamics 
\cite{painter_nature, tomita_prl, kippenberg_NatPhys, benson_apl2009, vahala_NatPhoto2010, takashima_apl, tanaka_optexp}.
Tapered-fiber-coupled microcavity systems thus occupy a central place in current cavity quantum electrodynamics research. 
In addition, tapered fibers are attractive sensing device that can efficiently capture the fluorescence from single light emitters \cite{nayak_optexp}, 
including quantum dots and biological molecules.
Recent theoretical studies have predicted that the fluorescence collection efficiency can be enhanced by up to 10 \% 
by employing a tapered fiber \cite{srinivasan_optexp}. 
This high collection efficiency is as high as that of a high numerical aperture objective lens. 

Despite these attractive applications, tapered fibers have a significant drawback; 
they rapidly lose their high transmittance soon after fabrication.
There have been several studies on this issue so far, which can be summarized as follows \cite{post-fabrication}. 
(1)  \textit{the degradation cannot be ascribed only to dust or particulate deposited on the surface}, 
even though this particle adsorption partly contributes to the degradation \cite{dustsuggest1, dustsuggest2}. 
(2) \textit{formation of cracks at the surface as a consequence of water absorption} 
is another possible cause \cite{JOpt_review}. 
(3) as a solution to this degradation, embedding the fiber in low-refractive-index polymer matrix 
was proposed and demonstrated \cite{post-fabrication, JOpt_review, jjap_Xu, IEICE_Vienne}.
These previous experiments were not performed under the condition 
where the dust-particle density and the humidity were precisely controlled.
In addition, embedding fibers in protecting matrix may not be useful for applications like micro-resonator couplings or sensing devices.

Here we present a systematic study of the transmittance degradation in tapered fibers by precisely controlling 
the dust-particle density and the humidity. 
The following conclusions we found based on our study are different from the previous ones. 
We have found that 
(1) the time degradation can be mostly ascribed only to the dust particles deposited on the surface and 
(2) that the effect of humidity is negligible to the transmittance degradation.
More importantly, by reducing the particle density down to cleanliness class 10, 
(3) we have succeeded in suppressing the degradation almost completely for 1 week with 850-nm-diameter tapered fiber 
and for more than 48 hours with thinner sample (450 nm in diameter), 
without using any post-fabrication treatment like embedding the 
fiber in polymer matrix.

\section{Experiments}

Tapered fibers were fabricated from standard single-mode optical fibers (Thorlabs, 630HP) 
by simultaneously heating and pulling them using a computer-controlled system equipped with a ceramic heater and a motorized stage \cite{konishi_apl}. 
This fabrication process was performed in a class-10 cleanroom (room A in Table \ref{table1}).
The transmittance in tapered fibers at $\lambda = 635$ nm was monitored during the fabrication process. 
The initial transmittance was defined as the ratio of the transmitted powers before and after the tapering.
We fabricated three tapered fibers with diameters ranging from 340 to 460 nm. 
All of them had the initial transmittance greater than 0.9 (see Table \ref{table1}). 
The length of the nanowire region, where the taper diameter was less than 1 $\mu$m, was typically 10 mm, 
and the length of the entire region of the tapering shape was approximately 30 mm.
These values of the initial transmittance, greater than 0.9, confirmed the near perfect adiabatic transitions in those tapered fibers. 
The tapered fibers were then immediately subjected to the following experiments. 

They were respectively stored in a cleanroom (room A), a semi cleanroom (room B), 
and a normal laboratory (room C); these rooms had different dust particle densities.
The dust densities in each room were monitored using a particle counter (Lighthouse, Handheld 3013). 
We detected three particle sizes: 0.3 $\mu$m, 0.5 $\mu$m, and 1.0 $\mu$m.
Table \ref{table1} summarizes the experimental conditions. 
The temporal profile of the transmittance of each tapered fiber was measured using the experimental setup 
shown in Fig. \ref{fig2}(a).
A red He--Ne laser ($\lambda = 632.8$ nm) was coupled to a single-mode optical fiber. 
The laser beam was split into two beams by a 50:50 fiber beam splitter. 
One beam was used as a reference and the other beam was coupled to the tapered fiber. 
The light passing through the tapered fiber was detected as the signal. 
The laser input power to the tapered fiber was $\sim$100 $\mu$W.
Laser intensity fluctuations were compensated for by taking a moving average of the data over 60 s.

Note that the transmittance shown in Figs. \ref{fig2}(b)--(d) included losses at the fiber connection 
inserted before the tapered fiber, which consisted of a FC/PC connection and a spliced point as shown in Fig. \ref{fig2}(a).
The total insertion loss at this connection was 0.6--2.5 dB, which varied connection by connection, 
but it was stable unless disconnected again. 
This insertion loss provided the different stating transmittances at time 0 in Figs. \ref{fig2}(b)--(d).
Furthermore, to eliminate the possibility of transmittance degradation before the measurement, 
%we immediately performed the transmittance measurement after the fabrication of the tapered fibers as described above.
%During that time, 
we protected the tapered fibers until the start of the measurement by storing it in a dust-free sealing box 
(the transmittance degradation is negligible in a dust-free environment as discussed later) that   
allowed us to finish all the connections and setup with isolating the fiber from any unnecessary exposure to dust particles. 
For these reasons, the differences of the starting transmittance at time 0 in Figs. \ref{fig2}(b)--(d) are only due to the insertion loss at the connector.

%\begin{table}[t!]
%\centering
%\begin{tabular}{ccccc}
%\hline
%  &Room A 	&Room B	&Room C 	&  \\
%\hline
%Particle size & & & \\
%0.3 $\mu$m					& 40					& 93710				& 550940			 \\
%0.5 $\mu$m 					& 9						& 6433				& 40330\\
%1.0 $\mu$m					& 1						& 57					& 1893	 \\
%& & \multicolumn{2}{l}{\ \ Unit: [$\rm{counts} \cdot \rm{ft}^{-3}$]}  \\
%\hline
%Cleanliness class & 10 & 10000 & N/A \\
%Humidity [\% ] & 42 & 30 & 36 \\
%\hline
%\hline
%Taper diameter [nm] &450				&340			&460		 \\
%Initial transmittance &0.95				&0.9			&0.9		 \\
%\hline
%\end{tabular}
%\caption{Experimental conditions for each tapered fiber. The cleanliness class follows the definition of US Federal Standard 209E. 
%The taper diameter was measured by scanning electron microscopy. The initial transmittance of the tapered fiber was measured at 635 nm.}
%\label{table1}
%\end{table}

\begin{table}[t!]
\centering
\begin{tabular}{ccccc}
%\multicolumn{2}{l}{Particle size $\rm{counts} / \rm{ft}^{3}$ & & &  \\
\hline
  &Room A 	&Room B	&Room C 	&  \\
\hline
%\multicolumn{2}{l}{Dust particle density [$\rm{counts} / \rm{ft}^{3}$]} & & & \\
Particle size & & & \\
0.3 $\mu$m					& 40					& 93710				& 550940			 \\
0.5 $\mu$m 					& 9						& 6433				& 40330\\
1.0 $\mu$m					& 1						& 57					& 1893	 \\
& & \multicolumn{2}{l}{\ \ Unit: [$\rm{counts} \cdot \rm{ft}^{-3}$]}  \\
\hline
Cleanliness class & 10 & 10000 & N/A \\
Humidity [\%] & 42 & 30 & 36 \\
\hline
\hline
%\multicolumn{2}{l}{Taper diameter [nm]} && & \\
Taper diameter [nm] &450				&340			&460		 \\
Initial transmittance &0.95				&0.9			&0.9		 \\
%& && Unit: [nm]  \\
\hline
\end{tabular}
\caption{Experimental conditions for each tapered fiber. The cleanliness class follows the definition of US Federal Standard 209E. 
The taper diameter was measured by scanning electron microscopy. The initial transmittance of the tapered fiber was measured at 635 nm.}
\label{table1}
\end{table}

\section{Results and Discussions}

\begin{figure}[t!]
\centering
	\includegraphics{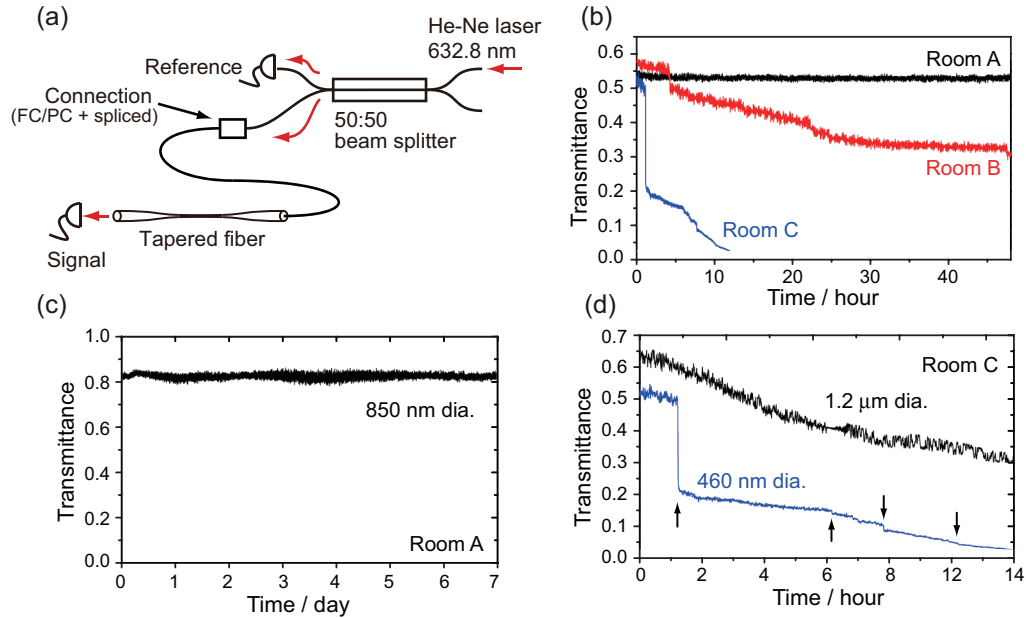}% Here is how to import EPS art
	\caption{(Color online) 
	(a) schematic diagram of the experimental setup used to measure the transmittance profile.
	(b) transmittance profiles of the three tapered fibers stored in a cleanroom (room A), 
	a semi cleanroom (room B), and a normal laboratory (room C).
	(c) the transmittance profile of the 850-nm-diameter tapered fiber placed in room A.
	(d) comparison of the transmittance degradation dynamics of thin (460 nm diameter) and thick (1.2 $\mu$m diameter) tapered fibers in room C.
	Arrows indicate sudden reductions in the transmittance.}
	\label{fig2}
\end{figure}

Figure \ref{fig2}(b) shows the temporal profiles of the transmittance of the three tapered fibers. 
It reveals that dust particles considerably affect the transmittance. 
The tapered fiber stored in room A exhibited very little degradation in its transmittance.
The degradation is less than 1\% of the starting transmittance (\textit{i.e.} at time 0) in 48 hours, which is less than the noise.
In contrast, the transmittance of the tapered fiber in room C deteriorated very rapidly over 10 h. 
In particular, at about 4 h the dust considerably reduced the optical transmittance ($T = 0.5 \to 0.2$).
The tapered fiber in room B exhibited intermediate behavior. 
Note that the tapered fiber in room B was smaller than the others (see Table \ref{table1}) and so may be more sensitive to dust particles. 
The semi-clean room B is therefore not as relatively bad as Fig. \ref{fig2}(b) may suggest.

It should be emphasized that transmittance was conserved as long as tapered fibers are stored in room A.
Figure \ref{fig2}(c) shows a transmittance profile of a 850-nm-diameter tapered fiber placed in room A. 
It indeed did not show any detectable degradation of the transmittance, \textit{e.g.} less than the noise ($\pm$0.02), over a week.
These results demonstrate that cleanroom conditions are very effective in preserving the transmittance of tapered fibers.

These temporal profiles provide insight into the degradation dynamics of the transmittance. 
The sudden drops in the transmittance observed at several points indicate that dust particles, which can significantly scatter the guided light, 
were adsorbed onto the surface of the taper region. 
Gregor \textit{et al.} found that the transmittance of the tapered fiber was instantaneously reduced by the deposition of an ionized fluorobead to the 
taper surface \cite{gregor_optexp}. 
They also performed a numerical simulation and found that magnitude of the transmittance reduction depends on the adsorbed particle size. 
It is therefore natural to assume that different-sized dust particles adsorbed on the tapered fiber and caused the transmittance reductions.

\begin{figure}[t!]
\centering
	\includegraphics{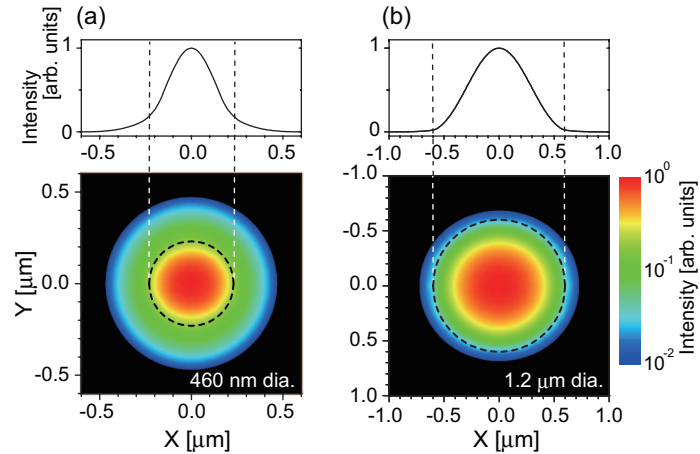}% Here is how to import EPS art
	\caption{(Color online) calculated spatial mode profiles of propagating light for (a) thin and (b) thick tapered fibers. 
	The upper panel is the intensity cross section in the plane $Y$ = 0.
	The lower panel shows a two-dimensional view of the propagating light. 
	The dashed line indicates the tapered fiber diameter.
	The black-color region is where the intensity is less than 0.01.
	A waveguide mode solver (Fimmwave, Photon Design) was used in these simulations. 
	%(c) Schematic illustrations for the dust adsorption to the thick and thin tapered fibers, which cause discrete drops in the transmittance profile.
	}
	\label{fig3}
\end{figure}

To understand the cause of these discrete drops in more detail, 
we placed a thick tapered fiber (1.2 $\mu$m in diameter) in room C and measured its transmittance profile. 
The result is shown in Fig. \ref{fig2}(d) together with data for the 460-nm-diameter tapered fiber already shown in Fig. \ref{fig2}(b). 
The transmittance of the thick tapered fiber deteriorated gradually without any sudden drops 
in contrast to the profile of the 460-nm-diameter tapered fiber.
The fact that the sudden drops were observed only in the thin tapered fiber can be explained 
by considering the evanescent field generated in the taper region. 
Figure \ref{fig3} shows simulated spatial mode profiles of light propagating ($\lambda$ = 633 nm) in the thin and thick tapered fibers. 
A large evanescent field was generated in the thin tapered fiber [Fig. \ref{fig3}(a)], 
whereas the light was almost completely confined in the silica cladding of the thick tapered fiber [Fig. \ref{fig3}(b)]. 
The evanescent field accounted for 9.5\% of the total intensity in the thin tapered fiber, while it was only 0.93\% of the total intensity in the thick fiber.
The range of the evanescent field around the thin fiber was 240 nm in the radial axis, 
which was three times longer than 80 nm of the range around the thick fiber. 
Here the range of the evanescent field was defined as the distance from the taper surface 
to where the intensity became 1\% of the maximum intensity in Fig. \ref{fig3}.
The large evanescent field generated around the thin tapered fiber is responsible for the high sensitivity of the transmittance to dust particles, 
namely for the sudden drops.
%
%Note that only the $HE_{11}$ mode was excited in all of the tapered fibers used in Fig. \ref{fig2}(b) because of the adiabatic tapering transition. 
%This fact excludes the possibility of the enhancement of the loss originated from the scattering of higher order modes that have 
%higher intensity at the surface than the $HE_{11}$ mode.

We next discuss the effects of oxidation and humidity on the transmittance degradation. 
First, oxidation evidently has negligible effect on the degradation since the transmittance was preserved in a clean atmosphere containing 20\% of oxygen.
Second, humidity might be an additional cause of the optical degradation as reported previously \cite{post-fabrication, JOpt_review, humidity_effect}.
We investigated the effect of humidity on the degradation by placing tapered fibers in room A (cleanroom) with various humidities 
ranging from 12--42\%, and no detectable degradation was observed.
In addition we placed a tapered fiber (400 nm in diameter) in a box filled with 86\%-humidity air in room A
and found that the transmittance was still conserved after 2 days within the range of the laser source fluctuation.
These results show that humid air does not deteriorate the optical transmittance.

It should be emphasized that this is the first report that demonstrates the conservation of the high optical transmittance 
in tapered fibers without any post-fabrication treatment such as embedding the fiber in polymer matrix.
One can naturally expect that omitting post-fabrication treatment is more preferable to the applications 
like micro-resonator coupling or sensing applications.
The present systematic study on the cause of the transmittance degradation has revealed 
that the degradation is principally ascribed to the dust particles deposited on the surface 
and it is not related to the humidity.
On the basis of these results, by placing the tapered fibers in a dust-free environment, 
we have demonstrated the preservation of the transmittance over 1 week with a negligible loss of less than the noise ($\pm$0.02).

\section{Summary}
The effect of dust particle density and humidity on the transmittance degradation of tapered fibers has been systematically studied.
The results have revealed that the degradation is due to dust particles. 
Water adsorption and oxidation were found not to affect the optical transmittance.
Storing fibers in a dust-free environment can preserve the optical transmittance
so that the degradation is less than the noise ($\pm$0.02) in 1 week.
The comparison of the degradation dynamics of the thick and thin tapered fibers, together with the waveguide simulations, 
has suggested that the large evanescent field generated around the thin tapered fiber 
is responsible for the high sensitivity of the transmittance to dust particles, \textit{i.e.} for the discrete drops in the transmittance. 
These findings are useful for tapered-fiber-based research involving 
quantum optics, white-light continuum generation, and particle nano-scale sensing.

\section*{Acknowledgements}
We would like to thank Prof. Hakuta for his suggestion on the 
importance of the clean environment for tapered fibers.
We gratefully acknowledge financial support from MIC-SCOPE, JST-CREST, 
JSPS Grant-in-Aid for Scientific Research on Innovative Areas ``Quantum Cybernetics" (No. 21102007), 
JSPS Grant-in-Aid for Scientific Research (Nos. 20244062 and 21840003), 
JSPS-FIRST, Special Coordination Funds for Promoting Science and Technology (MEXT), 
and the G-COE program.

\end{document}